\documentclass[letterpaper]{iopart}
\usepackage{iopams}
\jl{6}

\input epsf

\def\be {\begin{equation}}
\def\ee {\end{equation}  }
\def\beq{\begin{eqnarray}}
\def\eeq{\end{eqnarray}  }

\def\lp{\left(}
\def\rp{\right)}
\def\lb{\left[}
\def\rb{\right]}

\def\sq{^2}
\def\PD{\partial}
\def\rarrow{\rightarrow}
\def\tri{\triangle}

\def\rhor{\rho_o}
\def\bfq{{\bf q}}
\def\bff{{\bf f}}
\def\bfw{{\bf w}}
\def\bfs{{\bf s}}

\def\bfr{{\bf r}}

\def\bfF{{\bf F}}
\def\tbfq{\tilde{\bfq}}
\def\tbfw{\tilde{\bfw}}

\def\hbfq{\hat{\bfq}}
\def\hbff{\hat{\bff}}
\def\hbpsi{\hat{\bpsi}}
\def\bbfq{\bar{\bfq}}
\def\iph{{i+1/2}}
\def\imh{{i-1/2}}
\def\bfv{{\bf v}}

\def\JCP{{\it  J.\ Comput.\ Phys.\ }}

\def\mstrut{\rule{0mm}{4mm}}

\begin{document}

\title{Ultrarelativistic fluid dynamics}

\author{David W. Neilsen and Matthew W. Choptuik}
\address{Center for Relativity,
         The University of Texas at Austin,
         Austin, TX 78712-1081}

\begin{abstract}
This is the first of two papers examining the critical collapse
of spherically symmetric perfect fluids with the equation of state
$P = (\Gamma -1)\rho$.  Here we present the equations of motion and
describe a computer code capable of simulating the extremely relativistic
flows encountered in critical solutions for $\Gamma \le 2$.
The fluid equations are solved using a high-resolution shock-capturing scheme
based on a linearized Riemann solver.
\end{abstract}


\submitted

%\maketitle
\vskip 2pc

%%%%%%%%%%%%%%%%%%%%%%%% BEGIN PAPER %%%%%%%%%%%%%%%%%%%%%%%%%%
\section{Introduction}
\label{sec:introduction}

This paper describes a new computer code which simulates a self-gravitating,
relativistic perfect fluid in spherical symmetry.  The fluid model uses  
an equation of state $P=(\Gamma -1)\rho$, where $P$ and $\rho$ are 
the fluid pressure and total energy density, respectively, and $\Gamma$ is 
a constant satisfying $1 \le \Gamma \le 2$.  
The code has been optimized for {\em ultrarelativistic} fluid flows,
that is, for flows with  Lorentz factors much larger than unity. 
This optimization involves a novel definition of the fluid variables, the use of
a modern high-resolution shock-capturing scheme, and care in reconstruction of
the primitive fluid variables---such as pressure and velocity---from 
the conserved quantities which are actually evolved by the code.

Our new code was specifically developed to study the critical gravitational
collapse of perfect fluids.  Critical collapse has become an 
interesting subfield in general relativity since its initial
discovery in the massless Klein-Gordon system~\cite{mwc93},
and the perfect fluid model has played an important role
in advancing our understanding of the critical phenomena which arise 
at the threshold of black hole formation. 
(For an excellent introduction to critical phenomena, see the
review by Gundlach~\cite{cg97}.)
While the critical solutions for perfect fluids in spherical 
symmetry have been the subject
of recent study~\cite{cejc,kha,dm,hka,kha2,pbmc,goliath,carr1,carr2}, 
the precise nature of the critical solutions
for $\Gamma \gtrsim 1.89$ was not previously
known, and thus one of the chief goals of our investigation was a thorough
analysis of this regime.  In the remainder of this paper we 
describe the equations of motion which are solved, and the numerical 
techniques which we use to solve them.   A companion paper~\cite{dnmc2}
describes in detail the results we have generated with the code.

%%%%%%%%%%%%%%%%%%%%%%%%% SECTION %%%%%%%%%%%%%%%%%%%%%%%%%%%%%
\section{Geometry and fluid model}
\label{sec:intro_equations}

The Einstein equations couple the spacetime geometry, encoded
in the Einstein tensor, $G_{ab}$, to the stress-energy tensor, $T_{ab}$,
associated with the matter content of the spacetime: 
\be
G_{ab} = 8\pi T_{ab},
\ee
(here and throughout, we use units 
in which the speed of light, and Newton's gravitation constant
are unity: $c=1$ and $G=1$, and Latin indices $a, b, c, \cdots$ take on
the spacetime values $0, 1, 2, 3$.)
A fluid is a continuum model for a large number of
particles that uses macroscopic properties of a thermodynamic
system, such as internal energy and pressure, as fundamental 
dynamical variables. 
A perfect fluid has no shear stresses or dissipative forces,
and has a stress-energy tensor
\be
T_{ab} = (\rho + P)u_a u_b + Pg_{ab},
\ee
where $\rho$ is the energy density, $P$ is the pressure,
$u_a$ is the fluid's four-velocity, and $g_{ab}$ is the spacetime metric.
The energy density $\rho$
contains {\it all} contributions to the total energy, which for a perfect
fluid include the rest mass energy density, $\rhor$,
and the internal energy density
\be
\rho = \rhor + \rhor\epsilon,
\ee
where $\epsilon$ is the specific internal energy.
The fluid number density, $n$, is related to $\rhor$ via
\be
\rhor = mn,
\ee
where $m$ is the rest mass of a single fluid particle.
The basic equations of motion for the fluid can be derived from 
local conservation of (a) the energy-momentum 
\be
\nabla_aT^{ab} = 0,
\ee
and (b) the particle number
\be
\nabla_a \lp n u^a \rp = 0,
\ee
where $\nabla_a$ is the (covariant)
derivative operator compatible with $g_{ab}$.
To these conservation laws one must adjoin an 
equation of state, $P = P(\rhor,\epsilon)$, which, 
further, must be consistent with the first law of thermodynamics.

%-----------------------SUBSECTION----------------------------------------
\subsection{Equation of state}
\label{sec:eos}

The equation of state (EOS) closes the fluid equations by providing a
relationship between the pressure and (in our case) the rest energy density
and internal energy.  The nature of this relationship provides
much of the physics for a given system.
As mentioned in the introduction, our primary motivation for exploring 
ultrarelativistic fluid dynamics is 
to study perfect fluid critical solutions.
We expect these solutions to be scale invariant (self-similar),
and we therefore choose an EOS compatible with this symmetry.
The EOS
\be
P = (\Gamma - 1)\rho,
\label{eq:ureos}
\ee
where $\Gamma$ is a constant,
is the {\em only} EOS of the form $P=P(\rho)$ which 
is compatible with
self-similarity~\cite{mcat,aotp,ce93},
and is notable for the fact that it results in a sound speed, $c_s$, which
is independent of density:
\be
c_s = \sqrt{\frac{dP}{d\rho}}
    = \sqrt{\Gamma -1}.
\ee
One can argue that 
this EOS is particularly appropriate for ultrarelativistic 
fluids, and hence we will refer 
to \eref{eq:ureos} as the {\it ultrarelativistic equation
of state.}
We note that 
the EOS for a ``radiation fluid'' corresponds to $\Gamma = \frac{4}{3}$,
while $\Gamma=1$ gives a pressureless fluid (dust).
We do not consider the case of dust collapse here; hence, in
what follows, $1 < \Gamma \le 2$.

Another important fluid model is the ideal gas with the equation of state
\be
\label{eq:igeos}
P=(\Gamma -1)\rhor\epsilon.
\ee
In the ultrarelativistic 
limit, the kinetic energy of the constituent particles of the fluid
(or internal energy of the fluid in a thermodynamic context)
is much larger than the mass energy, $\rhor\epsilon \gg \rhor$,
giving $\rho\approx \rhor\epsilon$. 
Thus, one can interpret the EOS \eref{eq:ureos} as 
the ultrarelativistic limit of the ideal-gas EOS.  
As discussed in~\cite{dnmc2}, the ideal-gas EOS, in the 
ultrarelativistic limit, becomes, in a limiting sense, scale invariant.  
As the critical solutions reside in this
ultrarelativistic limit, the critical solutions for fluids
with the ideal-gas EOS are reasonably expected to be identical to 
the critical solutions computed using
\eref{eq:ureos}.  For this reason we hereafter limit our attention to 
the ultrarelativistic equation of state.

%-----------------------SUBSECTION----------------------------------------
\subsection{Geometric equations of motion}
\label{sec:geometric_eqs}

We use the ADM 3+1 formalism (specialized to spherical symmetry) to integrate 
the Einstein equations, and 
choose polar-areal coordinates
for simplicity of the equations of motion and for singularity avoidance.
Specifically, adopting a polar-spherical coordinate system
$(t,r,\theta,\phi)$, we write the spacetime metric as 
\be
\rmd s\sq = -\alpha(r,t)\sq\,\rmd t\sq + a(r,t)\sq\,\rmd r\sq 
        + r\sq\,\left( \rmd\theta\sq
        + \sin\sq\theta\,\rmd\phi\sq \right),
\label{eq:metric}
\ee
wherein the radial coordinate, $r$, directly measures proper surface area. 
In analogy with the usual Schwarzschild form of the static spherically symmetric 
metric, it is also useful to define the mass aspect function
\be
  m(r,t) \equiv \frac{r}{2}\lp 1 - \frac{1}{a\sq}\rp .
\ee
The fluid's coordinate velocity, $v$, and 
the associated Lorentz gamma function, $W$, are defined by
\be
v(r,t) \equiv \frac{au^r}{\alpha u^t}, \qquad
W(r,t) \equiv \alpha u^t.
\ee
Since the fluid four-velocity is a unit-length, time-like vector ($u^a u_a = -1$),
we then have the usual relation between $W$ and $v$:
\be
W\sq = \frac{1}{1 - v\sq} .
\ee
We now introduce two {\em conservation} variables
\beq
\label{eq:tsdef}
\tau(r,t) &\equiv  (\rho + P)W\sq - P \nonumber \\
S(r,t) &\equiv  (\rho + P)W\sq v,
\eeq
so named because they allow the fluid equations of motion to be 
written in {\em conservation form} (albeit with the addition of a source
term), as discussed in detail in \sref{sec:conservation}.  
In contrast to the conservation variables, we refer to the 
quantities $P$ and $v$ as {\em primitive} variables.
With the above definitions, the non-zero components of the
stress-energy tensor are  given by
%
%\be
%\eqalign{
%T^t{}_t &=  - \tau \\
%T^t{}_r &=  \displaystyle\frac{a}{\alpha}S\\
%T^r{}_r &=  Sv + P\\
%T^\theta{}_\theta &=  T^\phi{}_\phi =  P}
%\ee
%
\be
\begin{array}{lccl}
T^t{}_t =  - \tau  &&& T^r{}_r =  Sv + P \\
T^t{}_r =  \displaystyle\frac{a}{\alpha}S 
            &&& T^\theta{}_\theta =  T^\phi{}_\phi =  P.\\
\end{array}
\ee

A sufficient set of Einstein equations for the geometric variables $a$ and
$\alpha$ are given by (a) the non-trivial component of the momentum constraint 
(the notation $\PD_x f$ denotes partial
differentiation, i.e.\ $\PD_x f \equiv \PD f/\PD x$\/)
\be
%\frac{a_{,t}}{a} = - 4\pi r\alpha a S,
\PD_t a = - 4\pi r\alpha a^2 S,
\label{eq:momcon}
\ee
and by (b) the polar slicing condition, which follows from the demand that 
metric have the form~(\ref{eq:metric}) for all $t$:
\be
\label{eq:polar_slicing}
%(\ln\alpha)_{,r} = a^2 \lb 4\pi r\lp Sv + P \rp + \frac{m}{r^2} \rb.
\PD_r (\ln\alpha) = a^2 \lb 4\pi r\lp Sv + P \rp + \frac{m}{r^2} \rb.
\ee
An additional equation for $a(r,t)$,
\be
\label{eq:hamcon}
%\frac{a_{,r}}{a} = a^2 \lp 4\pi r\tau - \frac{m}{r^2} \rp.
\PD_r a = a^3 \lp 4\pi r\tau - \frac{m}{r^2} \rp \, ,
\ee
follows from the Hamiltonian constraint.

%-----------------------SUBSECTION----------------------------------------
\subsection{Fluid equations of motion}
\label{sec:fluid_eom}

Given the ultrarelativistic EOS~(\ref{eq:ureos}), the time evolution of 
our perfect fluid is completely determined by $\nabla_a T^{ab}=0$.
The derivation of the equations of motion---which can 
can naturally be written in conservation form---is a straightforward 
piece of analysis, and will not be given in detail here. 
Instead, we will simply quote the results, and for convenience in discussing 
the numerical method of solution, we adopt a 
``state vector'' notation. We thus define two-component vectors
$\hbfq$ and $\bfw$, which are the conservation 
and primitive variables, respectively
\be
\hbfq
    \equiv \lb  \begin{array}{cc} \tau \\  S \end{array} \rb\! , \qquad
\bfw \equiv \lb \begin{array}{cc} P \\ v \end{array} \rb \!.
\ee
We then define a ``flux vector,'' $\hbff$, and a ``source vector'' $\hbpsi$
\be
\hbff \equiv \lb\begin{array}{cc} S \\ Sv + P \end{array}\rb \qquad
\hbpsi \equiv \lb \begin{array}{cc} 0 \\ \Sigma \end{array} \rb\! .
\ee
These variables have been introduced with a hat $(\,\hat{}\,)$
to distinguish them from the new variables defined in \sref{sec:newvars},
which are subsequently used in the actual numerical solution algorithm.
Further, to expedite the discretization of the equations of motion, 
we decompose the source term, $\Sigma$, into two pieces, as follows:
\be
\Sigma \equiv \Theta + \frac{2\alpha P}{ar},
\ee
where
\be
\Theta \equiv (Sv - \tau)  \lp 8\pi\alpha a r P + \alpha a \frac{m}{r\sq}\rp
  + \alpha a P\frac{m}{r^2}.
\ee
We note that 
in spherically symmetric Minkowski spacetime we have $\Theta= 0$ and
$\Sigma = 2P/r$.
With the above definitions, we can now write the 
fluid equations of motion in the conservation form
\be
\PD_t \hbfq +
\frac{1}{r^2}\PD_r \lp r^2 X\hbff\rp = \hbpsi,
\label{eq:con_eom}
\ee
where
\be
X \equiv \frac{\alpha}{a}
\ee
is a purely geometric quantity.

Written in the above form, 
the fluid equations of motion \eref{eq:con_eom} contain a mixture of conservation 
and primitive variables, and thus it is necessary to transform
between both sets of variables at each step in the integration
procedure.
The primitive variables $\bfw$ can be expressed in terms of
the conservation variables 
$\hbfq$ by inverting the definitions~\eref{eq:tsdef}
of the conservation variables:
\be
P = -2\beta\tau + \lb 4\beta\sq\tau\sq
+ (\Gamma-1)(\tau\sq - S\sq)\rb^{\frac{1}{2}}
\label{eq:pdef}
\ee
\be
v = \frac{S}{\tau + P},
\label{eq:vdef}
\ee
where the non-negative constant $\beta$ is defined by
\be
\beta \equiv \frac{1}{4}\lp 2 - \Gamma \rp.
\ee
The pressure equation~(\ref{eq:pdef}) comes from the solution of 
a quadratic with a specific root 
chosen to yield a physical (non-negative) pressure.
This demand ($P \ge 0$)
further requires that $\tau \ge |S|$.  A second physical
requirement is that $v$ be bounded by the speed of light, $|v| \le 1$,
and from~(\ref{eq:vdef}) this will clearly be automatically satisfied when 
$\tau \ge |S|$.  
These physical restrictions on
the primitive variables can sometimes be violated in numerical
solutions of the fluid equations, and we discuss some numerical
techniques aimed at ameliorating such difficulties 
in sections \ref{sec:floor} and \ref{sec:velocity}.
Finally, we note that the above transformation from $\tbfq$ to $\bfw$
is particularly simple in that it can be expressed algebraically.  
The corresponding transformations for the gamma-law gas EOS~\eref{eq:igeos}
involves a transcendental equation which, in a numerical
implementation, must be solved iteratively at each grid point.

%-----------------------SUBSECTION----------------------------------------
\subsection{New conservative fluid variables}
\label{sec:newvars}

\begin{figure}
\epsfxsize = 10cm
\centerline{\epsffile{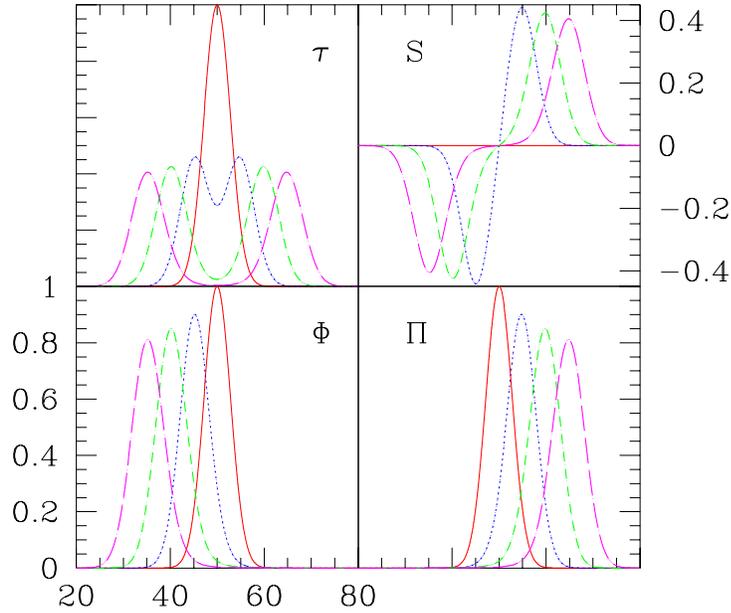}}
\caption{These plots show various fluid quantities at four different instances
(equally spaced in time)
in a flat spacetime, slab-symmetric evolution with $\Gamma = 1.9$.
The initial configuration is a time-symmetric
Gaussian pulse.  
The top frames show the evolution of the original conservation variables,
$\tau$ and $S$.  As the evolution proceeds, the pulse separates
into left and right-moving halves, and a vacuum region ($\tau \to 0$)
develops between the two sub-pulses.
The bottom frames show the evolution of the 
new conservation variables, $\Pi$ and $\Phi$, which are specifically 
defined so as to avoid the formation of such vacuum regions. 
The correspondence of the new variables to left and
right moving ``waves'' is also evident.  
Note that the plots of $\tau$, $\Pi$ and $\Phi$ have the same
vertical scale, while
the vertical scale for $S$ is shown separately.
The horizontal (radial) scale is the same for all of the plots.
}
\label{fig:piphi}
\end{figure}

Using the conservation variables $\hbfq$ defined above,
and the numerical method described in sections \ref{sec:numerical_methods}
and \ref{sec:solving}, we developed a preliminary code to solve the 
relativistic fluid equations.  We then tested this code by considering 
evolutions in Minkowski spacetime using slab and spherical symmetry.
The tests in slab symmetry were completely satisfactory, modulo the
convergence limitations of the numerical scheme. 
However, in spherical symmetry, we found that our method frequently failed
for ``stiffer'' fluids ($\Gamma \gtrsim 1.9$), most notably in ``evacuation 
regions'' where $\rho\rarrow 0$.  Additionally, the fluid in such
regions often became {\em extremely} relativistic, and the combination
of  $\rho \rightarrow 0$ and $|v| \rightarrow 1$ proved particularly 
difficult to simulate.
These problems that we encountered in spherical symmetry led us
to seek a new set of conservation variables, and to motivate
this change of variables, first consider the evolutions shown in \fref{fig:piphi}.
Here we begin with a time-symmetric, spherical shell of fluid,
which has a Gaussian energy density profile. 
Due to the time-symmetry, as the evolution unfolds,
the shell naturally splits into two sub-shells---one in-going
and one out-going---and as the sub-shells separate,
a new evacuation region forms in the region where the fluid was originally 
concentrated.  Examination of the conservation variable profiles reveals that
$|S|\approx \tau$, and this observation 
suggests that we adopt new variables
\be
\Phi \equiv \tau - S, \qquad
\Pi \equiv \tau + S,
\ee
which loosely represent the in-going ($\Phi$) 
and out-going ($\Pi$) parts of the solution.
Thus our new state vector of conservation variables is 
\be
\bfq \equiv \lb  \begin{array}{cc} \Pi \\  \Phi \end{array} \rb\! .
\ee
Not surprisingly, the numerical difficulties in evacuation regions 
are not completely cured with this change of variables;
however, the new variables $\bfq$ provide a {\em significant}
improvement over $\tbfq$ 
in evolutions of spherically symmetric fluids with $\Gamma \gtrsim 1.9$.

The equations of motion for the new variables $\bfq$
can be readily found by adding
and subtracting the two components of \eref{eq:con_eom}, giving 
\be
\PD_t \bfq + \frac{1}{r\sq}\PD_r \lb r\sq X \bff \rb = \bpsi,
\label{eq:eom_new_vars}
\ee
where the flux and source terms are now given by 
\be
\bff \equiv \lb \begin{array}{c}
\frac{1}{2}(\Pi - \Phi)(1+v) + P \\
\mstrut
\frac{1}{2}(\Pi - \Phi)(1-v) - P \end{array} \rb,
\qquad
\bpsi \equiv \lb\begin{array}{r} \Sigma\\
\mstrut -\Sigma\end{array} \rb\! .
\ee
The transformation from conservative to primitive variables can be found
by simply changing variables in \eref{eq:pdef} and \eref{eq:vdef}
\be
P = -\beta(\Pi + \Phi) + \lb \beta^2\lp \Pi + \Phi\rp^2
     + (\Gamma -1)\Pi\Phi\rb^\frac{1}{2},\\
\label{eq:ppiphi}
\ee
\be
v = \frac{\Pi - \Phi}{\Pi + \Phi + 2P}.
\label{eq:vpiphi}
\ee
We note that, given $\tau > |S|$, the new variables $\bfq$ are strictly 
positive: $\Pi > 0$, $\Phi > 0$.

%-----------------------SUBSECTION----------------------------------------
\subsection{The perfect fluid as a scalar field}
\label{sf}

There is a well-known relation between an irrotational, 
stiff ($\Gamma=2$) perfect fluid
and a massless Klein-Gordon scalar field.  In this section we discuss
the relationship between scalar fields and perfect fluids for
$0 < \Gamma \le 2$.
 The perfect fluid equations of motion 
\be
\nabla_aT^{ab} = 0,
\ee
can be written in terms of $\rho$, $P$, and $u^a$ as
\be
u^a \nabla_a + (\rho + P)\nabla_a u^a = 0,
\label{eq:I}
\ee
\be
(\rho + P)u^a \nabla_a u^b + \lp g^{ab} + u^a u^b\rp\nabla_a P = 0.
\label{eq:II}
\ee
If we assume the ultrarelativistic equation of state,
$P=(\Gamma-1)\rho$,
then these equations become
\be
\nabla \lp \rho^{1/\Gamma}u^a\rp = 0,
\ee
\be
u^a \nabla_a u^b 
   + \frac{(\Gamma - 1)}{\Gamma} \lp g^{ab} + u^a u^b\rp \nabla_a \ln\rho = 0.
\ee
We seek a specific combination of $\rho$ and $u^a$ that allows
the fluid equations to be written in terms of a single variable,
and therefore introduce the {\it ansatz}
\be
w^a \equiv \rho^\mu u^a,
\ee
where $\mu$ is a constant that will be determined below.  
From elementary contractions
we can express both $\rho$ and $u^a$ in terms of $w^a$
\be
\rho = \lp -w_a w^a \rp^{\frac{1}{2\mu}},
\label{eq:rho_w}
\ee
\be
u^a = \lp -w_b w^b\rp^{-\frac{1}{2}} w^a.
\label{eq:u_w}
\ee
However, it remains to see if $\mu$ can be chosen such that 
$w^a$ will satisfy the fluid equations of motion.
We substitute expressions \eref{eq:rho_w} and \eref{eq:u_w}
into the momentum equation \eref{eq:II}, and find that
this equation is satisfied provided that
\be
\mu = \frac{\Gamma-1}{\Gamma},
\ee
{\it and}
\be
\nabla_{[a} w_{b]} = 0.
\label{eq:w_circ}
\ee
This latter condition allows one to write $w^a$ as the gradient of a 
scalar field
\be
w_a = \nabla_a \varphi.
\ee
The equation of motion for $\varphi$
is obtained from \eref{eq:I}
\be
\nabla_a \lb \lp -\nabla_c \varphi \nabla^c\varphi \rp^\nu 
           \nabla^a\varphi\rb = 0,
\ee
where
\be
\nu = \frac{2 - \Gamma}{2(\Gamma - 1)}.
\ee

The condition \eref{eq:w_circ}, $\nabla_{[a}w_{b]}=0$, 
reduces to the requirement that the fluid be irrotational
\be
\nabla_{[a}u_{b]} = 0.
\ee
Thus, the fluid equations for an ultrarelativistic, irrotational
fluid can be written in 
terms of a nonlinear equation for a scalar field, $\varphi$.
For the stiff fluid ($\Gamma=2$), we find that the equation of motion
for $\varphi$ becomes the massless Klein-Gordon equation
\be
\nabla_a\nabla^a \varphi = 0.\qquad\qquad (\Gamma=2)
\ee

One typically places physically motivated conditions on the fluid
variables, such as $\rho > 0$ and $u^a u_a =-1$.   Solutions of
the Klein-Gordon equation, however, have time-like, null, and
space-like gradients ($\nabla_a\varphi$).  With the usual physical
constraints on the fluid, then only a subset of possible Klein-Gordon
solutions can be interpreted as $\Gamma=2$ perfect fluids, namely those with
$\nabla_a\varphi\nabla^a\varphi < 0$.

%%%%%%%%%%%%%%%%%%%%%%%%%%% SECTION %%%%%%%%%%%%%%%%%%%%%%%%%%%
\section{Numerical methods for fluid equations}
\label{sec:numerical_methods}

An important consideration for numerical solutions of
compressible fluid flow is how the numerical method will respond to 
the presence or formation of
shocks, i.e. discontinuities in the fluid variables.
These discontinuities often cause the
dramatic failure of na\"\i ve finite difference schemes, and as
shocks form {\em generically} from smooth initial data, many special 
techniques have been developed for the numerical solution
of fluid equations.
One approach is to introduce an {\em artificial viscosity} 
that adds extra dissipation in the vicinity of a shock, 
spreading the would-be-discontinuity over a few grid points.
This technique has been widely used, and has the advantages 
of simplicity of implementation and computational efficiency.
However, Norman and Winkler~\cite{nw} investigated the use of
artificial viscosity in relativistic flows, and showed that an {\em explicit}
numerical scheme treats the artificial viscosity term inconsistently
in relativistic fluid dynamics, leading to large
numerical errors in the ultrarelativistic limit, $W\gg 1$.
A second approach to solving the fluid equations with shocks
comes from methods developed specifically for conservation
laws.  These methods, usually variations or extensions of
Godunov's original idea~\cite{godunov}
to use piece-wise solution of the Riemann problem, 
have proven to be very reliable and robust.
LeVeque~\cite{rjlbook,rjl98} has written excellent 
introductions to conservative methods,
and our presentation here is in the spirit of his work.
Furthermore, 
the application of these methods to problems in relativistic astrophysics
has been recently reviewed by Ib\'a\~nez and Mart\'\i~\cite{ibanez_marti}.
However, for the sake of completeness, we first briefly define and 
discuss conservation laws, and outline a
general approach for their solution.  We then discuss a linear
Riemann solver and a cell reconstruction method that results in a
scheme which, for smooth flows, is second order accurate in the 
mesh spacing.

\subsection{Conservation methods}
\label{sec:conservation}

Conservation laws greatly simplify the mathematical description of
physical systems by focusing on quantities $\cal Q$---where $\cal Q$ may
be a state vector with multiple components---that do not change with time
\be
\PD_t \int_V \rmd {\cal Q} = 0.
\ee
In this section we discuss the derivation of numerical schemes
for this specific and important case where
$\int \rmd \cal Q$ is conserved on the computational domain.  
Our discussion will be general, and not specifically tailored for
the fluid PDEs derived in \sref{sec:newvars},
but for simplicity we restrict the discussion to one dimensional (in space)
systems.

While conservation laws are often written in 
{\em differential} form  (e.g.\ $\nabla_a T^{ab} = 0$) it is 
useful to first consider an {\em integral} formulation, which is
often the more fundamental expression.
Consider an arbitrary volume or cell, ${\cal C}_i$, with 
a domain $[x_1,x_2]$. 
The quantity of $\cal Q$ within ${\cal C}_i$ is denoted ${\cal Q}_i$, 
and we define a density function $\bfq$ such that
\be
{\cal Q}_i = \int_{x_1}^{x_2} \rmd x\, \bfq.
\label{eq:con_dens}
\ee
The change of ${\cal Q}_i$ with time can be calculated 
from the flux, $\bff(\bfq)$, of $\bfq$
through the cell boundaries.  
This consideration thus yields our conservation law:
\be
\frac{\rmd}{\rmd t} \int_{x_1}^{x_2} \rmd x \, \bfq(x,t)  
   = \bff(\bfq(x_1,t)) - \bff(\bfq(x_2,t)).
\label{eq:gen_con_law}
\ee
The conservation law
can be written in {\em integral} form by integrating
\eref{eq:gen_con_law} from an initial time, $t_1$, to a 
final time, $t_2$,
\beq
& &\int_{x_1}^{x_2} \rmd x \, \bfq(x,t_2) = \nonumber \\
& &\qquad \int_{x_1}^{x_2} \rmd x \, \bfq(x,t_1) 
  + \int_{t_1}^{t_2}\rmd t \, \bff(\bfq(x_1,t)) 
  - \int_{t_1}^{t_2}\rmd t \, \bff(\bfq(x_2,t))
\label{eq:con_law_int}
\eeq
and the differential form follows from further manipulation {\em if}
we assume that $\bfq$ is differentiable:
\be
\PD_t \bfq + \PD_x \bff(\bfq) = 0.
\label{eq:con_law_diff}
\ee
It should be emphasized that the integral formulation
should be viewed as {\em the} primary mathematical form 
for a conservation principle,
because it is {\em not} dependent on an assumption of differentiability.
For example, at a shock front in a fluid system, $\bfq$ is not
differentiable, and the differential form of the conservation law fails,
while the integral formulation is still satisfied.
Discretizations of conservation equations  via finite differences 
rely on the differential form, and artificial viscosity must be
added near shock fronts, forcing $\bfq$ to be differentiable.
An alternate strategy is to develop numerical algorithms
based directly on the integral formulation of the conservation laws.
The Godunov method and its extensions are examples of this latter approach,
and are the topic of the next section.

\subsection{Godunov's Method}
\label{sec:godunov}
 
Numerical algorithms for conservation laws are developed
by discretizing the equations in their fundamental integral form.
These methods derive from a {\em control volume} discretization, 
whereby the domain is divided into {\em computational} cells, $C_i$, now 
defined to span the interval $[x-\tri x/2,x+\tri x/2] \equiv [x_\imh,x_\iph]$, 
where $\tri x$
is the (local) spatial discretization scale.
Following the derivation of the integral conservation
law~\eref{eq:con_law_int} for the computation cell $C_i$,
we introduce the {\it averaged} quantities, $\bbfq_i^n$:
\be
\bbfq_i^n = \frac{1}{\tri x}\int_{x_{i-1/2}}^{x_{i+1/2}} \rmd x\, \bfq(x,t_n),
\label{eq:ave}
\ee
with $t_n \equiv n\tri t$, where $\tri t$ is the temporal
discretization scale.  We then obtain the discrete form
of the conservation law~\eref{eq:con_law_int}
\be
\bbfq^{n+1}_i = \bbfq^n_i -\frac{\tri t}{\tri x}
   \lp \bfF_{i+1/2} - \bfF_{i-1/2} \rp,
\label{eq:con_gen_method}
\ee
where the ``numerical flux'' is defined by
\be
\bfF_{i+1/2} = \frac{1}{\tri t}\int_{t_n}^{t_{n+1}} \rmd t \, 
    \bff(\bfq({x_{i+1/2}},t).
\label{eq:flux_int}
\ee

At first blush, a numerical method based on a discretization of the
integral conservation law does not appear promising: the flux 
integral~\eref{eq:flux_int} does not appear readily solvable,
and it generally is not.
However, in his seminal work, 
Godunov~\cite{godunov} devised a technique to approximately 
evaluate the flux integral by replacing the function $\bfq(x,t_n)$
with $\tbfq(x,t_n)$,
where $\tbfq(x,t_n)$ is a piece-wise constant function.
In this approach, the individual cells (``control volumes'') 
are treated as a sequence of ``shock tubes'', and a separate Riemann 
initial value problem is solved at each cell interface.  
Provided that the waves from neighboring cells do not interact---a proviso 
which gives a Courant-type condition on the time-step---each 
Riemann problem can be solved exactly to
yield the local solution $\tbfq(x,t)$ (for $t>t_n$) for each ``shock tube.''
Furthermore, since the solution of each of the local Riemann problems is 
self-similar,
$\tbfq(x_\iph,t)$ is a constant in time, and the evaluation of
the integral~\eref{eq:flux_int} becomes trivial.  
This then allows one to find explicit expressions for the cell averages 
at the advanced time,
$\bbfq^{n+1}$, via~\eref{eq:con_gen_method}.
In summary, the Godunov method proceeds as follows: 
(a) From the average $\bbfq_i^n$, one ``reconstructs'' a piece-wise constant
function $\tbfq(x,t_n)$ to approximate the solution in $C_i$;  
(b) the Riemann problem is solved at the interfaces between cells, giving
the solution $\tbfq(x,t)$ for $t_n < t \le t_{n+1}$;
(c) the solution $\tbfq(x,t_{n+1})$ is averaged over the cell $C_i$ to
obtain the average at the advanced time, $\bbfq^{n+1}_i$.
We note that methods for solving the Riemann problem exactly
for relativistic fluids have been given by Smoller and Temple~\cite{smoller}
for the ultrarelativistic EOS, and by 
Mart\'\i\ and M\"uller~\cite{mm} for the ideal-gas EOS.

Godunov's method has many nice properties: in particular, it 
is conservative
and allows for the stable evolution of strong shocks. 
However,  the original scheme {\em does} have some shortcomings:
convergence is only first order, and the exact solution of the Riemann 
problem may be computationally expensive, especially for relativistic
fluids.
% Dave, I'm commenting this out since it's essentially a round-about
% way of saying that the scheme is only first order
%Moreover, the complete Riemann solution contains many details
%that are ``thrown away'' in the Godunov procedure, because only 
%the single value 
%$\tbfq(x_\iph,t)$ is required to evaluate the integral~\eref{eq:flux_int}.
%(Equivalently, one could say that these features 
%in $\tbfq(x,t_{n+1})$ are ``washed out''
%in the averaging process~\eref{eq:ave} to obtain $\bbfq_i^{n+1}$.)
The convergence of the scheme can be improved by providing a
more sophisticated reconstruction $\tbfq(x,t_n)$, giving what are 
known as 
{\it high-resolution shock-capturing} methods.  One such procedure
is described in \sref{sec:cell_recon},  with details concerning
the scheme's convergence given in \sref{sec:tests}.
In order to address the issue of computational efficiency, approximate
Riemann solvers have been developed that relate the
problem-at-hand to a simpler system, for
which the Riemann problem is easier to solve.
Several approximate Riemann solvers have been developed for classical
fluid dynamics, and many of these approximate methods have been
extended to relativistic fluid systems.  These include
relativistic two-shock solvers~\cite{balsara,dai_woodward},
a relativistic HLLE solver~\cite{schneider},
and,
as discussed in \sref{sec:linear_riemann},
various linearized solvers.

%-----------------------SUBSECTION----------------------------------------
\subsection{Cell reconstruction}
\label{sec:cell_recon}

\begin{figure}
\epsfxsize = 8cm
\centerline{\epsffile{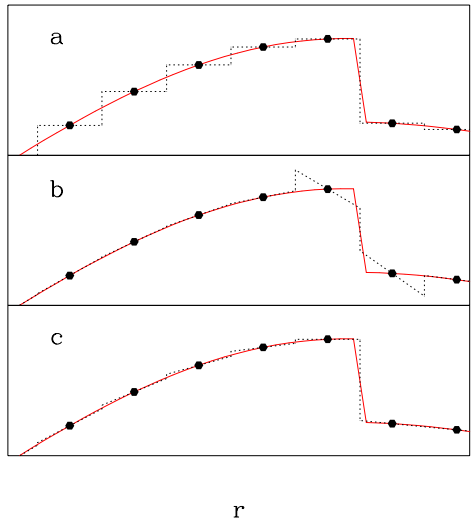}}
\caption{The three frames of this plot show different ways
a discretized function can be reconstructed in a control-volume
numerical method.
The solid line represents a continuum (or ``analytic'') function
and the solid hexagons represent discrete, approximate values of the function 
defined at grid points.  Frame (a) represents
the piece-wise constant reconstruction.  Frame (b) shows a na\"\i ve
piece-wise linear reconstruction of each cell using ${\bf s}_{i+1/2}$.  
This reconstruction oscillates
near discontinuities in the function---such oscillations can easily lead 
to instabilities.  Frame (c) shows a piece-wise linear reconstruction performed
with the minmod limiter as described in the text. This reconstruction 
produces a discrete representation of the dynamical variable which 
remains well-behaved near discontinuities.}
\label{fig:lim}
\end{figure}

Godunov-type numerical methods are based on solutions of the
Riemann initial value problem at the interfaces between cells.
As discussed above, during an update step one introduces 
functions $\tbfq(x,t)$--- defined piece-wise on the 
intervals $[x_{i-1/2},x_{i+1/2}]$---to approximate the solution in the 
control volumes $C_i$.  
These functions are created
from the cell averages $\bbfq^n_i$, and hence are called 
{\em reconstructions}.
Consider the cell interface at $x_\iph$: the state of the 
fluid immediately to the
right (left) is $\tbfq^r_\iph$ ($\tbfq^\ell_\iph$).
The simplest reconstruction is to assume that $\tbfq$ is piece-wise constant
\be
\tbfq^\ell_\iph = \bbfq_i,  \qquad
\tbfq^r_\iph = \bbfq_{i+1},
\ee
as used in the original Godunov method and, as already discussed, 
this reconstruction results in a numerical scheme in which
the spatial derivatives (and hence the overall scheme) 
have first order accuracy.  The convergence
can be improved by using a higher-order reconstruction for $\tbfq$,
but care must be exercised so that the reconstruction does not induce
spurious oscillations near discontinuities (see \fref{fig:lim}).

We have chosen to use a piece-wise {\it linear} reconstruction for $\tbfq$,
which formally results in a scheme with second order convergence.  
(The convergence properties are discussed in greater detail in
\sref{sec:tests}.)  The $\tbfq$ are reconstructed using
the total variation diminishing (TVD) minmod limiter introduced by
van Leer \cite{vl}.  
The van Leer limiter forces $\tbfq$ to be monotonic near discontinuities,
and this reduces the (local) accuracy of the scheme to first order.
The first step of the reconstruction algorithm
involves the computation of the slope (derivative
of the dynamical variable) centered at the cell boundaries
\be
\bfs_{i+1/2} = \frac{\bbfq_{i+1} - \bbfq_{i}}{r_{i+1} - r_{i}}.
\ee
A ``limited slope'', $\bsigma_i$, is then calculated via
\be
\bsigma_i =  \hbox{minmod}(\bfs_{i-1/2},\bfs_{i+1/2}),
\ee
where the minmod limiter is defined by
\be
{\rm minmod}(a,b) = \cases{
    a & if $|a| < |b|$ and $ab>0$\\
    b & if $|b| < |a|$ and $ab > 0$\\ 
    0 & if $ab < 0$.}
\ee
Using the limited slopes, we evaluate $\bbfq$ at the cell interfaces
as follows:
\be
\tbfq^\ell_{i+1/2} = \bbfq_i + \bsigma_i(r_{i+1/2} - r_i)
\ee
and
\be
\tbfq^r_\iph = \bbfq_{i+1} + \bsigma_{i+1}(r_{i+1/2} - r_{i+1}).
\ee
Finally, if we are unable to calculate physical values for 
$\tbfw^\ell$ and $\tbfw^r$ (a situation which {\em can} and {\em does}
occur owing to the finite-precision nature of our computations)
we revert to a piece-wise
constant reconstruction for $\tbfq^\ell$ and $\tbfq^r$.

%-----------------------SUBSECTION----------------------------------------
\subsection{The Roe linearized solver}
\label{sec:linear_riemann}

Perhaps the most popular approximate Riemann solver is the linearized
solver introduced by Roe \cite{roe}.
This solver (and subsequent variants)
has been used in a variety of applications involving general relativistic
fluids~\cite{ibanez,eulderink,romero,banyuls,brandt,font}, 
and has proven to be robust and efficient.  (The efficiency comparison
is relative to solving either the exact Riemann problem for relativistic
fluids, or a nonlinear approximation, such as the two-shock solver.)
As the name suggests,
the linearized solver approximates the full nonlinear problem by replacing the
nonlinear equations by {\it linear} systems
defined at each cell interface.  The associated 
linear Riemann problems can then be solved exactly and cheaply, and the 
resulting solutions can be pieced together to produce an approximation
to the solution of the original, nonlinear equations.  Thus, in 
order to understand the 
Roe scheme, it is instructive to first consider linear conservation laws. 

The linear, scalar advection equation
\be
\PD_t q + \lambda\PD_x q = 0,
\ee
has the well-known solution
$q(x,t) = q(x-\lambda t,0),$
where $\lambda$ is a  constant and $q(x,0)$ specifies the initial state.
This scalar solution can be extended to linear systems of
conservation equations
\be
\PD_t\bfq + A \PD_x \bfq = 0,
\label{eq:lin_sys}
\ee
where $A$, an $M\times M$ {\it constant} matrix, is, by
assumption, diagonalizable, with
real eigenvalues, $\lambda_\mu$.  
(Greek indices take the values $1, \ldots , M$.) 
Let $R$ be the matrix of right eigenvectors, $\bfr_\mu$, of $A$:
\be
R \equiv [\bfr_1|\ldots|\bfr_M],
\ee
and let $\Lambda$ be the diagonal matrix:
\be
\Lambda \equiv \hbox{diag}[\lambda_1,\ldots,\lambda_M].
\ee
We then have
\be
A = R\Lambda R^{-1},
\ee
and the solution of the system may be  obtained by introducing
``characteristic variables'', $\bfv$:
\be
 \bfv = R^{-1}\bfq.
\ee
Using characteristic variables, the equations \eref{eq:lin_sys} decouple
into a set of scalar advection equations
\be
\PD_t \bfv + \Lambda\PD_x\,\bfv = 0,
\ee
which can be immediately solved via:
\be
	\bfv_\mu(x,t) = \bfv_\mu(x - \lambda_\mu t,0).
\ee
Given $\bfv(x,t)$,
the transformation $\bfq = R \bfv$ then produces the solution 
of~~\eref{eq:lin_sys} in terms of the original variables, $\bfq$.
%Using this general solution method, the Riemann problem for the
%linear system~\eref{eq:lin_sys} can be easily obtained~\cite{rjlbook}.

Turning now to the nonlinear case, the key idea is to first write the 
nonlinear system in {\em quasilinear} form
\be
\PD_t \bfq + {\cal A}(\bfq)\,\PD_x \bfq = 0.
\label{eq:lincons}
\ee
Here, $\cal A$ is an $M\times M$ matrix which is now a function of $\bfq$.
Roe~\cite{roe} gives three specific criteria for the construction of $\cal A$:
\begin{enumerate}
\item ${\cal A}(\tbfq^\ell,\tbfq^r)\lp \tbfq^r - \tbfq^\ell\rp = 
        \bff(\tbfq^r) - \bff(\tbfq^\ell)$;
\item ${\cal A}(\tbfq^\ell,\tbfq^r)$ is diagonalizable with real eigenvalues;
\item ${\cal A}(\tbfq^\ell,\tbfq^r) \to \bff'(\bfq)$ smoothly as
       $\tbfq^\ell,\tbfq^r \to \bfq$.
\end{enumerate}
The latter two criteria can generally be satisfied by letting $\cal A$ be
the Jacobian matrix evaluated using the arithmetic average of the
conservation variables at the interface:
\be
{\cal A} = \frac{\PD \bff(\bfq)}{\PD \bfq}\bigg\vert_{\bfq=\bbfq_\iph},
\label{eq:jacobian}
\ee
where
\be
\bbfq_\iph = \frac{1}{2}\lp \tbfq^\ell_\iph + \tbfq^r_\iph \rp.
\ee
While this construction does not generally satisfy the first criterion,
(\ref{eq:jacobian}) is often used 
in relativistic fluid dynamics~(see for example~\cite{ibanez,romero,font}) 
on the basis of its relative simplicity, and we also adopt this 
approach. 
On the other hand, other authors~\cite{eulderink} have constructed a 
linearized Riemann solver for relativistic fluids with true Roe averaging, and 
we therefore refer to our scheme as a ``quasi-Roe'' method.  

Having defined a specific linearization, the scheme proceeds by 
evaluation of ${\cal A}(\bbfq_\iph)$---which is now viewed as a 
matrix with (piecewise) constant coefficients---followed by the solution 
of the Riemann problem for the resulting
linear system.
Carrying through an analysis not given here~(see e.g.~\cite{rjlbook}), 
the Roe flux can be defined as 
\be
\bfF_{i+1/2} = \frac 1 2 \lb \bff(\tbfq^r_\iph) + \bff(\tbfq^\ell_\iph)
    - \sum_\mu |\lambda_\mu|\tri\bomega_\mu \bfr_\mu \rb.
\label{eq:roeflux}
\ee
where, again, $\lambda_\mu$ and $\bfr_\mu$ are the eigenvalues and 
(right) eigenvectors,
respectively, of ${\cal A}(\bfq_\iph)$.
The quantities $\tri\bomega_\mu$ are defined in terms of the 
the jumps in the fluid variables across the interface
\be
\tbfq^r_\iph - \tbfq^\ell_\iph = \sum_\mu \tri\bomega_\mu \bfr_\mu .
\ee
For completeness, 
we give explicit expressions for the eigenvectors and eigenvalues of 
the ultrarelativistic fluid system~\eref{eq:eom_new_vars} in \ref{sec:eigen}.  

Finally, it is important to remember that approximate Riemann
solvers produce approximate solutions, which, under certain conditions,
may diverge from the physical solutions.  
For example, 
concentrating on the Roe solver,
Quirk~\cite{quirk} has recently reviewed
several ``subtle flaws'' in approximate solvers.
Fortunately, the approximate solvers often fail
in different ways, and where one solver produces an
unphysical solution, another solver may give the physical solution.
Thus, it may be necessary to investigate a particular problem with
multiple approximate Riemann solvers.  Therefore, we have also
implemented Marquina's solver~\cite{marquina}, an alternative linear solver
that has also found application in relativistic fluid 
studies~\cite{donat,font}, as an option in our code.  
In addition to using the quasi-Roe and Marquina 
solvers to investigate the critical collapse of perfect fluids,
we also implemented the HLLE solver in an independent code.
We found that the quasi-Roe solver gave accurate solutions, and
 provided the best combination 
of resolution and efficiency for the critical collapse problem.
Consequently, the results presented in \cite{dnmc2} 
were obtained with this solver.

%%%%%%%%%%%%%%%%%%%%%%%%%%% SECTION %%%%%%%%%%%%%%%%%%%%%%%%%%%
\section{Solving the Einstein/fluid system}
\label{sec:solving}

This section deals with some details of our numerical
solution of the coupled Einstein/fluid equations, including
the incorporation of source terms into our conservation laws, 
regularity and boundary conditions,
and methods for calculating physical values for $\bfw$ in
the ultrarelativistic regime.
In addition, we describe the initial data and mesh structure 
we have used in our studies of critical phenomena in fluid 
collapse.  Finally, we conclude the section with some remarks 
on how we have tested and validated our code.

%-----------------------SUBSECTION----------------------------------------
\subsection{Time integration}
\label{sec:time}

In \sref{sec:newvars}
the fluid equations of motion were written  essentially in
conservation form, except that a source term, $\bpsi$, 
had to be included.
While this source term clearly 
breaks the strict conservation form of the equations, it can
be self-consistently incorporated into our numerical scheme by
using the method of lines to discretize space and time
separately.  
Specifically, 
the discretized fluid equations become
\be
\frac{d\bbfq_i}{dt} = -\frac{1}{r^2_i\tri r}
\lb \lp r^2 X \bfF\rp_{i+1/2}
- \lp r^2 X \bfF\rp_{i-1/2}\rb + \bpsi(\bbfq_i),
\label{eq:update}
\ee
where $\bbfq_i$ is the cellular average of $\bfq$, $\bfF_{i\pm1/2}$ 
are the numerical
fluxes defined by~(\ref{eq:roeflux}), and $X=\alpha/a$, as previously.
These equations can be integrated in time using standard techniques
for ODEs.  In particular, Shu and Osher~\cite{shuosher} have investigated 
different
ODE integration methods, and have found that the modified Euler method
(or Huen's method) is the optimal second-order scheme 
consistent with the Courant condition required for a stabile evolution.
We briefly digress to define this scheme for a general set of 
differential equations of the form
\be
\frac{\rmd\bfq}{\rmd t} = L(\bfq),
\ee
where $L$ is a spatial differential operator.
Let $\bfq^n$ be the {\it discretized} solution at time $t=n\tri t$, 
and $\hat L$ be the discretized differential operator.
The modified Euler method
is a predictor-corrector method, with predictor 
\be
\bfq^* = \bfq^n + \tri t\, \hat L(\bfq^n),
\ee
and corrector 
\be
\bfq^{n+1} = \frac{1}{2}(\bfq^n + \bfq^*) + \frac{1}{2}\tri t\,
\hat L(\bfq^*).
\ee
Again, we note that $\tri t$ is subject to a Courant (CFL) condition,
 which can be deduced empirically or possibly from a linearized stability 
analysis.

Particularly in comparison to the treatment of the fluid equations, 
numerical solution of the equations governing the geometric 
quantities $\alpha$ and $a$ is straightforward.
As discussed previously, 
the lapse, $\alpha$, is fixed by the 
polar slicing 
condition \eref{eq:polar_slicing},
while $a$ can be found from either the Hamiltonian 
\eref{eq:hamcon} or momentum \eref{eq:momcon} constraints.
We have used discrete, second-order, versions of both equations for 
$a$, and have obtained satisfactory results in both cases (the polar 
slicing equation is likewise solved using a second-order scheme.)
In general, however, (and 
particularly on vector machines) solution via the momentum 
constraint yields a far more efficient scheme, and we thus generally use 
the momentum equation to update $a$. 

Full
details of our numerical scheme are
presented in \ref{sec:implementation}.

%-----------------------SUBSECTION----------------------------------------
\subsection{Regularity and boundary conditions}
\label{sec:regbound}

In the polar-areal coordinate system, the lapse ``collapses'' exponentially
near an apparent horizon, preventing the $t=\,$constant surfaces from
intersecting the physical singularity which must develop interior to 
a black hole. As the slices ``avoid'' the singularity,
elementary flatness holds at the origin for all times in the
evolution, giving
\be
a(0,t) = 1.
\ee
At each instant of time,
the polar-slicing condition~(\ref{eq:polar_slicing}) determines the 
lapse only up to an overall multiplicative constant, reflecting the 
reparameterization invariance, $t \to {\tilde t}(t)$, of the polar
slices.
We chose to normalize the lapse function so that as $r\to\infty$,
coordinate time corresponds to  proper time.
On a finite computational domain, and provided no matter out-fluxes 
from the domain, this condition is approximated via
\be
\alpha a\Big|_{r_{\rm{max}}} = 1.
\ee

In spherical symmetry the fluid flows along radial lines, and given
that there are no sources or sinks at the origin, we have that
$v(0,t) = S(0,t) =  0$. Thus
\be
\Pi(0,t) = \Phi(0,t) = \tau(0,t).
\ee 
Regularity at the origin further require $\tau, \Pi$ and $\Phi$ to 
have even expansions in $r$ as $r\to 0$:
\be
	\tau(r,t) = \tau_0(t) + r^2 \tau_2(t) + \Or(r^4)
\ee
\be
	\Pi(r,t) = \Pi_0(t) + r^2 \Pi_2(t) + \Or(r^4) = 
              \tau_0(t) + r^2 \Pi_2(t) + \Or(r^4)
\ee
\be
	\Phi(r,t) = \Phi_0(t) + r^2 \Phi_2(t) + \Or(r^4) = 
               \tau_0(t) + r^2 \Phi_2(t) + \Or(r^4)
\label{eq:Phi_expansion}
\ee
On our radial grid $r_i, \,\, i = 1, 2, \cdots N$, 
we use these expansions to compute grid-function values defined at 
$r=r_1=0$ in terms of 
values defined at $r=r_2$ and $r=r_3$.  
Specifically, once the values $\Phi_2$ and 
$\Phi_3$ have been updated via the equations of motion, we compute 
$\Phi_1$ using a ``quadratic fit'' based on the 
expansion~(\ref{eq:Phi_expansion}):
\be
\Phi_1 = \frac{\Phi_2 r_3{}^2 - \Phi_3 r_2{}^2}%
{r_3{}^2 - r_2{}^2}.
\ee
We then set $\Pi_1 = \Phi_1$.

At the outer boundary we apply out-flow boundary conditions, which in
our case are simply first-order extrapolations for $\Pi$ and $\Phi$:
\be
\Phi_N = \Phi_{N-1}\qquad\Pi_N = \Pi_{N-1}.
\ee
In addition, two ghost cells ($r=r_{N+1}, r=r_{N+2}$)
are added at the outer edge of the grid for ease in coding the
cellular reconstruction algorithm~\cite{rjl98}.
These ghost cells are also updated with first-order extrapolation.

%-----------------------SUBSECTION----------------------------------------
\subsection{Floor}
\label{sec:floor}

The fluid model
is a continuum approximation, and, at least na\"\i vely, the fluid equations
become singular as $\rho \rightarrow 0$.
In these evacuation regions, both
the momentum and mass density are very small, and therefore the 
velocity---which loosely speaking is the quotient of the 
two---is prone to fractionally large numerical errors.
These errors then often result in the computation of unphysical
values for the fluid variables, such as supraluminal velocities,
negative pressures or negative energies. 
(In addition, of course, our code must contend with the usual 
discretization and round-off errors common to any numerical solution 
of a set of PDEs.)  At least from the point of view of Eulerian
fluid dynamics, it seems fair to say that a completely satisfactory 
resolution of the evacuation problem does not exist.  In the 
absence of a mathematically rigorous and physically acceptable 
procedure, we adopt the {\em ad hoc} approach of demanding that
$\rho > 0$ everywhere
on the computational domain, i.e.\ we exclude the possibility
that vacuum regions can form on the grid.
In terms of our conservation
variables $\bfq$, this requirement becomes $\Pi > 0$ and $\Phi > 0$.
In a wide variety of situations, our numerical solutions of the fluid
equations naturally satisfy these constraints.
However, the critical
solutions for ``stiff'' equations of state ($\Gamma \gtrsim 1.9$)
develop extremely relativistic velocities ($W > 10^6$) in regions
where $\rho$ is small~\cite{dnmc2},  and we are unable to solve
the fluid PDEs in these cases without imposing
{\em floor} (or minimum) values on $\bfq$.  Specifically, at each step 
in the integration we require
\be
\Pi \ge \delta,\quad \Phi \ge \delta,
\ee
where the floor $\delta$ is chosen to be several orders of 
magnitude smaller than the density associated with what we 
feel are the physically relevant
features of the solution---a typical value is $\delta = 10^{-10}$.
The floor is often applied in regions where $\Pi$ and $\Phi$ differ
greatly in magnitude, and discretization errors can easily lead to
the calculation of a negative value for either function.  
For example, the floor may be applied to the ``in-going'' function
in a region where the fluid is overwhelmingly ``out-going.''
In these cases, the effect of the 
floor is dynamically unimportant.  However, the floor may be
invoked in other cases, where its effect on the dynamics is less certain.

Given the {\em ad hoc} nature of this regularization procedure, 
the crucial question
is whether the floor affects the computed solutions in a substantial way.  
We investigated this question by comparing critical solutions
for $\Gamma=2$ (the most extreme case) which were
calculated with two distinct floor values:
$\delta = 10^{-8}$, and our usual $\delta = 10^{-10}$.
The two solutions appeared identical,
and identical mass-scaling exponents~\cite{dnmc2} were calculated.
However, we note that the use of a floor makes estimates of the 
maximum Lorentz factor
attained in the critical solutions 
unreliable because the largest velocities occur in regions where
the floor is enforced.

%-----------------------SUBSECTION----------------------------------------
\subsection{Calculating the velocity}
\label{sec:velocity}

The simple expression~\eref{eq:vdef} for $v$ in terms of $\bfq$,
when used na\"\i vely with finite precision arithmetic, 
can result in the computation of unphysical, supraluminal
velocities.
For example, when searching for critical solutions we routinely
calculate fluid flows with $W\gtrsim 10^3$.
Thus, when calculating $v$ from
the quotient~\eref{eq:vdef}, small numerical errors can easily
conspire to give $|v|>1$, rather than the correct $|v|\gtrsim 0.999999$.
On the other hand, the combination
\be
\chi \equiv W^2 v
\ee
is insensitive to small numerical errors, and provides a better
avenue for calculating $v$ from the conservation variables.
From the definition \eref{eq:tsdef} of $S$ we have
\be
\chi = \frac{(\Gamma - 1)}{\Gamma}\frac{S}{P}.
\ee
The velocity can then be calculated from $\chi$ using
\be
\label{eq:vchi}
v = \frac{1}{2\chi}\lp\sqrt{1 + 4\chi\sq} - 1\rp .
\ee
To the limit of machine
precision, $v$ is then in the physical range $-1 < v < 1$.
When $\chi \ll 1$, we calculate $v$ from a Taylor expansion of \eref{eq:vchi},
although \eref{eq:vdef} could also be used.
We also use $\chi$ when calculating $\bfw$ from $\bfq$ for the 
ideal-gas EOS~\eref{eq:igeos}.

%\be
%v\approx\chi\lp 1 + \chi\sq \lp 2\chi\sq - 1\rp\rp. \qquad\qquad (\chi \ll 1)
%\ee

%-----------------------SUBSECTION----------------------------------------
\subsection{Grid}
\label{sec:grid}

\begin{figure}
\epsfxsize = 7.5cm
\centerline{\epsffile{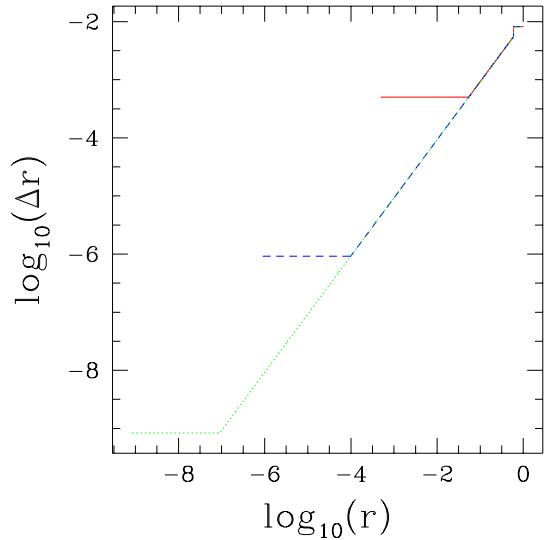}}
\caption{Illustration of the re-meshing algorithm used in 
investigations of critical collapse.  The grid spacing 
$\tri r$ is shown as a function of $r$ on
a log-log plot.
The solid line represents the initial grid, the dotted line shows
the grid spacing after the first addition of points near the origin,
and
the dashed line shows the grid spacing after the second regridding.  
Note that the grid spacings near the origin, and near the outer 
edge of the computational domain are uniform (horizontal lines).
At each regridding cycle, the grid spacing near
the origin is halved, and the new points are 
matched smoothly onto the previous grid.
A critical evolution may involve more than 20 regriddings, although only
a small number of points (50--150) may be added at a time.
}
\label{fig:grid}
\end{figure}

The black-hole-threshold critical solutions---which are our primary focus---are 
generically self-similar, and as such, require
essentially unbounded dynamical range for accurate simulation.
Thus some sort of adaptivity in the construction
of the computational domain is crucial, and, indeed, the 
earliest studies of critical collapse~\cite{mwc93} used 
Berger-Oliger adaptive mesh refinement~\cite{bergoliger} to great 
advantage.
However, in contrast to the early work, we know (at least schematically) 
the character of the critical solutions we seek, and thus we can, 
and have, used 
this information to construct a simple, yet effective, adaptive grid method.  
(Our approach is similar in spirit to 
that adopted by Garfinkle~\cite{garfinkle} in his study of scalar field 
collapse.)
Specifically, at any time during the integration our spatial grid has 
three distinct domains: the two regions near $r=0$ and $r=r_{\rm max}$ have
uniform grid spacings (but the spacing near $r=0$ is typically
{\em much} smaller than that near the outer edge of the computational
domain), and the intermediate region has grid points distributed uniformly
in $\log(r)$ (see  \fref{fig:grid}).
As a near-critical solution propagates to smaller spatial scales, 
additional grid points are added in order to maintain some given
number of grid points between $r=0$ and some identifiable
feature of the critical solution.  For example, we typically  require that at
least 300 or so grid points
lie between the origin and the maximum of the profile of the 
metric function $a$. 

The primary advantage of this gridding scheme is that it 
is simple to implement, and yet allows us to resolve detail over 
many length scales: the ratio of the grid
spacing at the outer edge to the spacing at the origin is typically
$10^{10}$--$10^{13}$ at the end of an evolution.
The primary disadvantage of this scheme is that it is 
specialized for critical collapse, and cannot be used for more general
physical problems. 

%-----------------------SUBSECTION----------------------------------------

\subsection{Initial data for critical solutions}

We expect that the critical solutions in fluid collapse
will be universal, in the sense that {\em any} family of initial
data which generates families that ``interpolate'' between 
complete dispersal and black hole formation, should exhibit the 
same solution at the black hole threshold.   
We have thus focused attention
on a specific form of initial data, which generates
initially imploding (or imploding/exploding) shells of fluid.  
Specifically, the energy density
in the shells has a Gaussian profile,
\be
\tau = \tau_o \exp\lb -(r - r_o)^2 / \Delta^2\rb + K,
\ee
where the constant $K$---typically of magnitude $10^{-6}\tau_o$---represents 
a constant ``background''.
It should be note that
this background is used only in setting the
initial data, and is not held fixed during the evolution---in particular
$K$ is {\it not} a floor as discussed in \sref{sec:floor}.
The shells are either time-symmetric, or have an initial inward velocity
which is proportional to $r$.  Critical solutions were found by
fixing $r_o$ and $\Delta$, and then tuning the pulse amplitude $\tau_o$.

%-----------------------SUBSECTION----------------------------------------

\subsection{Tests}
\label{sec:tests}

\begin{figure}
\epsfxsize = 10cm
\centerline{\epsffile{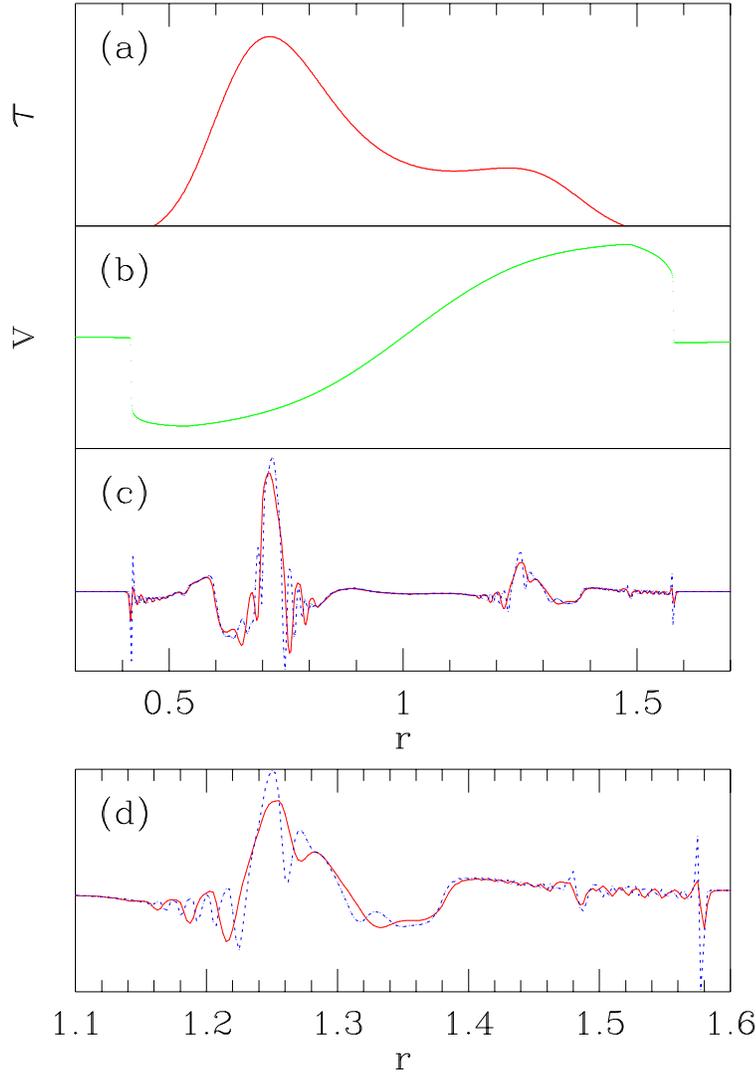}}
\caption{
Illustration of some of 
the convergence
properties of the solution algorithm discussed in the text.  
Here 
we evolved a time-symmetric shell of
fluid ($\Gamma = 1.3$)
using uniform grids with three different resolutions: $\tri r = h, 2h$ and
$4h$.
Convergence is investigated by
comparing the solutions obtained using  the three distinct discretization
scales.
In frame~(c), the solid line is $\lp\tau_{2h} - \tau_{4h}\rp$
and the dotted line is $4\lp\tau_h - \tau_{2h}\rp$, 
where the subscript on $\tau$ indicates the grid spacing
for a particular solution. 
When the convergence is second order, the two lines should
(roughly) coincide, while when the convergence is first order, the amplitude of
the dotted line should be twice that of the solid line.
As expected, we see that the convergence is not second order at the
shock.  (Of course the whole notion of convergence at a discontinuity
fails, as  the notion of Richardson expansion requires smooth 
functions.) However,
we also can see that the convergence is only first order at the
extrema of $\bfq$---at these points, the slope changes sign, and the
minmod limiter produces a first-order reconstruction.  Frame~(d) shows 
a more detailed view of a portion of the data displayed in (c).
For context, we also show $\tau$ in frame~(a) and $v$ in frame~(b). 
}
\label{fig:conv3}
\end{figure}

When developing a code such as the one described here, there are 
a number of tests which should be 
passed in order to provide confidence that the algorithm is 
producing reliable results.
Perhaps most fundamental of these is the convergence test, which 
generally demonstrates that the numerical method is consistent and 
has been correctly implemented, but which also provides an
{\em intrinsic} method for estimating the level of error in 
a given numerical solution.
For our high-resolution shock-capturing scheme,
a general rule-of-thumb is that the convergence 
should be (apparently) second order where the flow is
smooth, and first order at discontinuities,
 where the effects of the slope limiter become important.
In addition, we can also expect first order convergence
near extrema of $\bbfq$, since  at these points, the slope, $\bfs$,
changes sign, and the minmod limiter gives a piece-wise constant
reconstruction for $\tbfq$.
A convergence test where these
effects are apparent is shown in \fref{fig:conv3}.

After the numerical algorithm has been correctly implemented,
one often compares results from the code to known closed-form 
solutions.  
In the early stages of code development, we tested
the shock-capturing algorithm in this fashion by
solving initial data for a shock tube, and comparing the
results with the known solution of the Riemann problem.
While the shock-tube provides a good
test of the fluid solver, the test is done in Minkowski space with
slab symmetry,
and can probe neither the implementation of the geometric factors in the
fluid equations, nor the discretized Einstein equations.
A few general relativistic fluid systems {\em can} be solved exactly,
and have traditionally been used to test new codes.  These include
static, spherical stars (Tolman-Oppenheimer-Volkoff), spherical
dust collapse (Oppenheimer-Synder) and ``cosmological'' tests with
a Robertson-Walker metric.  In our companion paper~\cite{dnmc2}, we advocate
the use of perfect fluid critical solutions as an additional
test problem; one which involves both dynamic gravitational
fields and highly relativistic fluid flows.  
Thus we consider the ultimate test of our full GR/fluid 
code to be the dynamical calculation of self-similar 
perfect fluid critical solutions, which can then be compared to 
solutions computed directly (but also numerically!) from
a self-similar {\em ansatz}~\cite{dnmc2}.

%%%%%%%%%%%%%%%%%%%%%%%%%%% SECTION %%%%%%%%%%%%%%%%%%%%%%%%%%%
%\section{Conclusion}
%\label{sec:conclusion}

%We have presented the equations of motion for a spherically
%symmetric perfect fluid in general
%relativity, and have described a robust computer code which
%is capable of simulating 
%ultrarelativistic fluid flows, such as those encountered 
%in critical collapse.
%Results from this code, including comparisons with precisely 
%critical solutions generated from a self-similar {\em ansatz}
%are presented in a second paper.
%second paper.

%%%%%%%%%%%%%%%%%%%%%%%%%%% SECTION %%%%%%%%%%%%%%%%%%%%%%%%%%%
\appendix
\section{Characteristic structure}
\label{sec:eigen}

In this appendix we calculate the Jacobian matrix $\cal A$
for the relativistic fluid equations, and then compute the 
associated 
eigenvalues and right eigenvectors. 
The flat-space components of $\cal A$ are
\beq
{\cal A}_{11} &=& \frac{1}{2}\lp 1 + 2v - v^2\rp
                     + (1 - v^2)\frac{\PD P}{\PD \Pi}\\
{\cal A}_{12} &=& -\frac{1}{2}\lp v + 1 \rp^2
                     + (1 - v^2)\frac{\PD P}{\PD \Phi}\\
{\cal A}_{21} &=& \frac{1}{2}\lp v - 1\rp^2
                     + (v^2 - 1)\frac{\PD P}{\PD \Pi}\\
{\cal A}_{22} &=&  \frac{1}{2}\lp -1 + 2v + v^2 \rp
                     + (v^2 - 1)\frac{\PD P}{\PD \Phi},
\eeq
and the partial derivatives of $P$ are easily found from
\eref{eq:ppiphi}.
The eigenvalues $\lambda_\pm$ of $\cal A$
are the two roots of the quadratic equation
\be
\lambda_2 - ({\cal A}_{11} + {\cal A}_{22})\lambda + \det{\cal A} = 0,
\ee
and the right eigenvectors are
\be
\bfr_\pm = \lp \begin{array}{c} 1 \\ Y_\pm \end{array}\rp,
\ee
where
\be
Y_\pm \equiv \frac{\lambda_\pm - {\cal A}_{11}}{{\cal A}_{12}}.
\ee
If the eigenvalues are numerically degenerate owing to the limitations
of finite precision arithmetic, we set $\lambda_{\pm} = 0$.
When $\Gamma=2$, the eigenvalues and eigenvectors become simply
\be
\lambda_\pm = \pm 1, \quad
\bfr_+ = \lp \begin{array}{c} 1 \\ 0 \end{array}\rp , \quad
\bfr_- = \lp \begin{array}{c} 0 \\ 1 \end{array}\rp . \qquad\qquad (\Gamma =2)
\ee

%%%%%%%%%%%%%%%%%%%%%%%%%%% SECTION %%%%%%%%%%%%%%%%%%%%%%%%%%%
\section{Implementation Details}
\label{sec:implementation}

The origin in spherical symmetry requires additional care
because powers of $1/r$ appear in the flux and source terms.  One particular
difficulty results from the partial cancellation of the 
source term, $2\alpha P/(ar)$, with 
the pressure term in the flux.  Numerically this
cancellation is not exact, and this non-cancellation can induce
large errors near the origin. We therefore modify the difference equations
in order to eliminate the offending 
term.  We first decompose the numerical flux into
two parts $\bff^{(1)}$ and $\bff^{(2)}$:
\be
\bff^{(1)} = \lb \begin{array}{c}\frac{1}{2}(\Pi - \Phi)(1+v) \\
\mstrut
\frac{1}{2}(\Pi - \Phi)(1-v)\end{array} \rb  \qquad
\bff^{(2)} =  \lb \begin{array}{r} P \\
\mstrut
 -P \end{array} \rb,
\ee
so that $\bff = \bff^{(1)} + \bff^{(2)}$.
We then rewrite the conservation  equations \eref{eq:con_eom} with
these new fluxes as
\be
\PD_t \bfq +
\frac{1}{r^2}\PD_r \lp r^2 X \bff^{(1)}\rp 
+ \PD_r \lp X \bff^{(2)}\rp= \hat{\bSigma},
\ee
where the new source term $\hat{\bSigma}$ is
\be
\hat{\bSigma} =  \lb \begin{array}{r} \Theta \\ -\Theta \end{array} \rb.
\ee
The numerical flux function $\bfF$ is similarly decomposed: $\bfF = \bfF^{(1)}
+ \bfF^{(2)}$, with
\be
\bfF^{(1)}_{i+1/2} =
     \frac{1}{2} \lb \bff^{(1)}(\tbfq^\ell_{i+1/2})
 + \bff^{(1)}(\tbfq^r_{i+1/2}) 
      - \sum_\mu |\lambda_\mu|\tri\bomega_\mu \bfr_\mu \rb,
\ee
\be
\bfF^{(2)}_{i+1/2}= \frac{1}{2} \lb \bff^{(2)}(\tbfq^\ell_{i+1/2}) 
                 + \bff^{(2)}(\tbfq^r_{i+1/2}) \rb.
\ee

The finite-differencing of the flux terms is adapted so that the
derivatives have the correct leading order behavior near the origin.
From the regularity conditions discussed in~\sref{sec:regbound}
we have
\be
\lim_{r\rightarrow 0} r^2 X \bff^{(1)}\propto r^3, \qquad
\lim_{r\rightarrow 0} X \bff^{(2)}\propto  \rm{constant},
\ee
and we thus write the discretized equations of motion as
\beq
\frac{\rmd\bbfq_i}{\rmd t} & = & - \frac{3 \lb \lp r^2 X
  \bfF^{(1)}\rp_{i+1/2}
 - \lp r^2 X \bfF^{(1)}\rp_{i-1/2} \rb}%
{r^3_{i+1/2} - r^3_{i-1/2}} \nonumber\\
&  & \qquad \mbox{} -
\frac{\lp X\bfF^{(2)}\rp_{i+1/2} - \lp X\bfF^{(2)}\rp_{i-1/2}}%
{r_{i+1/2} - r_{i-1/2}}  +  {\hat{\bSigma}}_i.
\label{eq:actualeom}
\eeq

The geometric equations are differenced using standard second-order
finite-difference techniques.
The momentum 
constraint is
\be
\frac{\rmd a_i}{\rmd t} = 2\pi r_i\alpha_i a_i^2\lp \Pi_i - \Phi_i \rp,
\ee
and is integrated using the modified Euler method
described in \sref{sec:time}.
The discretized polar slicing condition~\eref{eq:polar_slicing} 
in discrete form is
\beq
& &(\ln\alpha)_{i+1}^n = (\ln\alpha)_i^n \nonumber \\
& & \qquad \mbox{} + \tri r \left\{ a
   \lb 2\pi r ((\Pi - \Phi)v + P) + \frac{1}{2r}\lp 1 
       - \frac{1}{a^2}\rp \rb\right\}_\iph^n\! ,
\eeq
where all of the basic variables---$a, \Pi, \Phi, v$ and $P$---in the  
$\{\}$ braces are evaluated at $r_\iph$ using an arithmetic average.

Finally, the overall flow of an integration step is as follows:
\begin{enumerate}

\item Begin with the data for time $t=t^n$:
  $\{\Pi^n,\Phi^n,P^n,v^n,\alpha^n, a^n\}$.

\item Reconstruct cells for $\{\tbfq^\ell,\tbfq^r\}$,
          and calculate the numerical fluxes $\bfF(\tbfq^\ell,\tbfq^r)$.

\item Perform the predictor step of the modified Euler method, obtaining
  $\{\Pi^*,\Phi^*,a^*\}$, then calculate $\{P^*,v^*\}$, and
  integrate the slicing condition to determine $\alpha^*$.

\item Reconstruct cells for $\{\tbfq^{\ell*},\tbfq^{r*}\}$, 
  and calculate the numerical
  fluxes $\bfF(\tbfq^{\ell*},\tbfq^{r*})$.

\item Perform the corrector step of the modified Euler method, obtaining
  $\{\Pi^{n+1},\Phi^{n+1},a^{n+1}\}$, then calculate $\{P^{n+1},v^{n+1}\}$, 
  and integrate the slicing condition to determine $\alpha^{n+1}$.

\item Check the regridding criteria, and regrid if necessary.

\end{enumerate}

\section*{Acknowledgments}
\label{sec:ack}
This work was supported in part by the National Science
Foundation under Grants PHY93-18152 (ARPA supplemented), PHY94-07194, 
PHY97-22068, by a Texas
Advanced Research Project grant, and by an NPACI award of computer time.
We thank Carsten Gundlach for interesting discussions, and for suggestions
on improving a preliminary version of this paper.
MWC gratefully acknowledges the 
hospitality of the Institute for Theoretical Physics, UC Santa Barbara, 
where part of this research was carried out.

\section*{References}
\begin{thebibliography}{99}

\bibitem{mwc93}   Choptuik M W 1993  \PRL {\bf 70} 9

\bibitem{cg97}    Gundlach C 1998 {\it Adv.\ Theor.\ Math.\ Phys.} {\bf 2} 1;
                  {\tt gr-qc/9712084}

\bibitem{cejc}    Evans C R and Coleman J S 1994 \PRL {\bf 72} 1782

\bibitem{kha}     Koike T, Hara T and Adachi S 1995 \PRL {\bf 74} 5170

\bibitem{dm}      Maison D 1996 \PL B {\bf 366} 82

\bibitem{hka}     Hara T, Koike T, and Adachi S 1996 {\tt gr-qc/9607010}

\bibitem{kha2}    Koike T, Hara T, and Adachi S 1999 \PR D {\bf 59}

\bibitem{pbmc}    Brady P R and Cai M J 1998 {\tt gr-qc/9812071}

\bibitem{goliath} Goliath M, Nilsson U S and Uggla C 1998 \CQG {\bf 15} 2841

\bibitem{carr1}   Carr B J, Coley A A, Goliath M, Nilsson U S and Uggla C
                  1999 {\tt gr-qc/990131}

\bibitem{carr2}   Carr B J, Coley A A, Goliath M, Nilsson U S and Uggla C
                  1999 {\tt gr-qc/990270}

\bibitem{dnmc2}   Neilsen D W and Choptuik M W 1999 submitted to \CQG;
                   {\tt gr-qc/9812053}

\bibitem{mcat}    Cahill M E and  Taub A H 1971 {\it Commun.\ Math.\ Phys.}
                          {\bf 21} 1

\bibitem{aotp}    Ori A and Piran T 1990 \PR D {\bf 42} 1068

\bibitem{ce93}    Evans C R 1993 in
                     {\it Lecture Notes of the Numerical Relativity
                     Conference, Penn State University, 1993} (unpublished).

\bibitem{nw}      Norman M L and Winkler K-H A 1986 {\it Astrophysical
                  Radiation Hydrodynamics} edited by Norman M L and
                  Winkler K-H A (Reidel, Dordrecht) p~449

\bibitem{godunov} Godunov S K 1959 {\it Mat. Sb.} {\bf 47} 271

\bibitem{rjlbook} LeVeque R J 1992 {\it Numerical Methods for Conservation
                          Laws} (Birkha\"user-Verlag, Basel)

\bibitem{rjl98}   LeVeque R J 1998  in {\it Computational Methods for
         Astrophysical Fluid Flow,} 27th Saas-Fee Advanced Course Lecture
         Notes (Springer-Verlag, Berlin, to be published); also available at
         {\tt http://sirrah.astro.unibas.ch/saas-fee/}.
%        and {\tt http://amath.washington.edu/\~rjl/}.

\bibitem{ibanez_marti} Ib\'a\~nez J M$^{\rm {\underline a}}$ and 
                       Mart\'\i\ J M$^{\rm {\underline a}}$ 
                       1999 preprint

\bibitem{smoller} Smoller J and Temple B 1993 {\it Commun. Math. Phys.}
                         {\bf 156} 67

\bibitem{mm}     Mart\'\i\ J M$^{\rm {\underline a}}$ and
                   M\"uller E 1994 {\it. J. Fluid Mech.} {\bf 258} 317

\bibitem{balsara} Balsara D S 1994 \JCP {\bf 114} 284

\bibitem{dai_woodward} Dai W and Woodward P R 1997 
              {\it SIAM J.\ Sci.\ Comput.\ } {\bf 18} 982

\bibitem{schneider} Schneider V, Katscher U, Rischke D H, Waldhauser B,
              Maruhn J A and Munz C-D 1993 \JCP {\bf 105} 92

\bibitem{vl}       van Leer B 1979 {\it J. Comp. Phys.} {\bf 32} 101

\bibitem{roe} Roe P L 1981 {\it  J.\ Comput.\ Phys.\ } {\bf 43} 357

\bibitem{ibanez} Ib\'a\~nez J M$^{\rm {\underline a}}$,
                 Mart\'\i\  J M$^{\rm {\underline a}}$,
                 Miralles J A and
                 Romero J V 1992 in {\it Approaches to Numerical Relativity,}
                (Cambridge University Press, Cambridge) p~223

\bibitem{eulderink}Eulderink F and Mellema G 1995 
                {\it Astronomy and Astrophysics Suppl.} {\bf 110} 587

\bibitem{romero} Romero J V,
                Ib\'a\~nez J M$^{\rm {\underline a}}$,
                Mart\'\i\  J M$^{\rm {\underline a}}$
                and  Miralles J A 1996
                {\it ApJ} {\bf 462} 839

\bibitem{banyuls}  Banyuls F, Font J A, 
                   Ib\'a\~nez J M$^{\rm {\underline a}}$,
                   Mart\'\i\  J M$^{\rm {\underline a}}$
                   and  Miralles J A
                   1997 {\it ApJ}  {\bf 476} 221

\bibitem{brandt}   Brandt S, Font J A, Ib\'a\~nez J M$^{\rm {\underline a}}$,
                   Mass\'o J and Seidel E 1998 {\tt gr-qc/9807017}

\bibitem{font}     Font J A, Miller M, Suen W and Tobias M 1998
                   {\tt gr-qc/9811015}

\bibitem{quirk}    Quirk J J 1994 {\it Int.\ J.\ Numer.\  Methods Fluids} 
                      {\bf 18} 555

\bibitem{marquina} Donat R and Marquina A 1996 {\it J.\ Comput.\ Phys.\ }
                         {\bf 125} 42

\bibitem{donat} Donat R, Font J A, Ib\'a\~nez J M$^{\rm {\underline a}}$
                   and Marquina A 1998 \JCP {\bf 146} 58

\bibitem{shuosher} Shu C-W and Osher S 1988 {\it J. Comput. Phys.}
                           {\bf 77} 439

\bibitem{bergoliger} Berger M J and Oliger J 1984 {\it J. Comput. Phys.}
                        {\bf 53} 484

\bibitem{garfinkle} Garfinkle D 1995 \PR D {\bf 51} 5558

\endbib

\end{document}